\documentclass[aps,amsmath,amssymb,twocolumn,prb,superscriptaddress]{revtex4-2}

\usepackage{graphicx}
\usepackage{dcolumn}
\usepackage{bm}
\usepackage{color}
\usepackage[normalem]{ulem}
\usepackage{amsmath}
\usepackage{hyperref}

\begin{document}

\title{The role of charge in thermodynamic uncertainty relations}

\author{David Christian Ohnmacht}
\email{david.ohnmacht@uni-konstanz.de}
\affiliation{Fachbereich Physik, Universität Konstanz, D-78457 Konstanz, Germany}

\author{Wolfgang Belzig} 
\affiliation{Fachbereich Physik, Universität Konstanz, D-78457 Konstanz, Germany}

\author{Juan Carlos Cuevas}
\affiliation{Departamento de F\'{\i}sica Te\'{o}rica de la Materia Condensada,
Universidad Aut\'{o}noma de Madrid, E-28049 Madrid, Spain}
\affiliation{Condensed Matter Physics Center (IFIMAC), Universidad Aut\'{o}noma de Madrid, E-28049 Madrid, Spain}

\date{\today}

\begin{abstract}
We demonstrate that the charge value of transport mechanisms heavily impacts the validity of thermodynamic uncertainty relations (TURs). Specifically, we show within the framework of full counting statistics, that the recently established quantum TUR can be violated by the presence of transport processes that carry more than one charge, like Andreev reflection processes in normal metal-superconductor junctions. We propose a modified quantum TUR, which incorporates the charge value and demonstrate that this charge-dependent quantum TUR can only be violated if the highest charge transport process exceeds this charge value. In particular, we establish that the breaking of the quantum TUR solely originates from the charge value of the highest charge transport process. Namely, our analytical considerations do not invoke the existence of superconductivity, and these considerations generally hold for non-interacting electronic transport which can be described by the scattering formalism.
\end{abstract}

\maketitle

\textit{Introduction} --- The classical thermodynamic uncertainty relation (TUR) establishes a trade-off relation between signal-to-noise ratio and 
the associated entropy production \cite{Barato2015,Todd2016}. In the case of charge transport, the TUR dictates $S/ I^2\geq 2k_{\rm B}/\dot\Sigma$ with $k_{\rm B}$ being the Boltzmann 
constant, $\dot\Sigma$ the entropy production, $I$ the average charge current and $S$ the corresponding noise \cite{Barato2015,Todd2016,Horowitz2020,Pietz2016,Pietzonka16b,Pietz2018,Timp2019,PhysRevB.98.155438,PhysRevResearch.5.043041}. TUR violations have been predicted in quantum systems attributed to either quantum coherence or the breaking of local detailed balance \cite{Andrieux_2007,PhysRevLett.120.090601,PhysRevX.11.021013,
PhysRevResearch.5.043041,PhysRevResearch.5.013038,PhysRevResearch.5.023155,PhysRevB.28.1655,Kheradsoud2019}. Large deviations of TURs due to quantum effect could enhance performance and stability of thermal machines \cite{PhysRevB.98.085425,PhysRevB.98.155438,PhysRevLett.120.090601,PhysRevE.99.062141}. There has been an ongoing effort on finding quantum systems which exhibit significant breakdowns of the TUR
\cite{PhysRevResearch.5.043041,PhysRevResearch.5.023155,PhysRevB.104.045424,PhysRevResearch.5.013038,PhysRevResearch.5.023155,PhysRevB.108.115422}, for example in hybrid normal-superconductor quantum dots \cite{PhysRevResearch.5.043041}. Whereas the classical TUR is the primarily studied TUR, there are other types of TURs, which arise from making different assumptions about the underlying system for example for driven systems \cite{Potanina2021,Menczel2021} or multiterminal systems \cite{Brandner2018}.

The classical TUR was studied for transport in quantum conductors in Ref.~\cite{PhysRevB.98.155438}, where a nonlinearity in the form of a energy dependent transmission function is needed in order to break the classical TUR. However, the breakings which originate from these non-nonlinearities are difficult to experimentally verify \cite{PhysRevB.101.195423}. It was shown in Ref.~\cite{Ohnmacht} that charge transport processes of charge values larger than one favor classical TUR violations, namely (multiple) Andreev reflection processes in superconducting junctions, which makes such systems promising candidates for the experimental demonstration of the breaking of the classical TUR. However, breaking the classical TUR does not quantify the severity of the violation. Namely, largely breaking the classical TUR is possible for a chain of quantum dots with a rectangular transmission, which corresponds to the maximal violation \cite{PhysRevB.104.045424}. A new tighter limit was established in Ref.~\cite{quantumTUR}, the so-called \textit{quantum TUR}, which marks the absolute limit for conventional conductors following Landauer's current formula \cite{Buttiker1986}. As shown in Ref.~\cite{natcomm}, this quantum TUR is broken in normal metal-superconductor junctions in the large gap limit, where Andreev reflection processes are dominating the transport, and the reason for the breaking of the quantum TUR was attributed to macroscopic quantum coherence arising from superconductivity. As transport in the presence of superconductors does not follow the assumptions made for the quantum TUR in Ref.~\cite{quantumTUR}, a violation is not fundamentally prohibited. Therefore, the question arises what the fundamental reason for the breaking of the quantum TUR is and if a new bound can be found for transport in hybrid junctions that also includes superconducting elements. Such a modified TUR has already been derived in Ref.~\cite{natcomm} for the large gap limit where quasiparticle tunneling is not present.

In this work we establish that the charge value of the dominating transport mechanism is the primary factor in determining the respective charge-dependent quantum TUR. We extend the quantum TUR to arbitrary charge values via the general framework of full counting statistics \cite{Levitov1993,Nazarov1999}. We demonstrate our findings by analyzing transport in normal metal-superconducting junctions numerically and show that the quantum TUR is easily broken but that transport obeys the \textit{charge-2 quantum TUR}. We furthermore show that transport in superconductor-superconductor junctions breaks \textit{any} charge quantum TUR because of the presence of an unbounded number of multiple Andreev reflection processes. We infer that there is no suitable charge quantum TUR for transport in superconductor-superconductor junctions and that the $n$th order multiple Andreev reflection in the limit $n\to \infty$ describes the absolute limit of TURs in quantum transport. Furthermore, the analytical considerations are illustrated in the framework of full counting statistics, without invoking the need of superconductivity. Therefore, we ultimately show that the breaking of the quantum TUR is solely attributed to charge values larger than one of a transport process and that the presence of superconductivity is merely the cause of this charge value to be present. 

\textit{Quantum TUR} --- In the following, we establish that transport processes which transfer more than one charge can break the quantum TUR. First, the quantum TUR establishes a trade-off between entropy production and noise to current ratio and can be written as \cite{quantumTUR}
\begin{equation}\label{Eq1}
    \sinh(\beta V/2) \frac{S}{I} \geq 1,
\end{equation}
with the inverse temperature $\beta = 1/(k_{\rm B}T)$, the voltage $V$, the noise $S$ and the current $I$ (we work in units of $e = 1$). Following Ref.~\cite{PhysRevB.98.155438,Ohnmacht}, we can perform perturbative expansions in the voltage of current and noise
\begin{align}
    I &= G_1V + \frac{1}{2} G_2 V^2 + \frac{1}{6} G_3 V^3 + ... \\
    S &= S_0 + S_1V + \frac{1}{2}S_2 V^2+ \frac{1}{6} S_3 V^3 + ...,
\end{align}
and of the prefactor
\begin{equation}
    \sinh(\beta V/2) =  (\beta V)/2 +(\beta^3V^3)/48 +...,
\end{equation}
with which we define the function
\begin{align}
    f_\alpha = (\beta V)/2 + \alpha \left[(\beta^3V^3)/48 + ...\right],
\end{align}
where we notice that for $\alpha = 0$, we retrieve the classical TUR \cite{PhysRevB.98.155438}, whereas we recover the quantum TUR for $\alpha = 1$. Then, we can expand Eq.~\ref{Eq1} in the voltage while using the relations $S_0 = 2G_1/\beta$ and $S_1 = G_2/\beta$ to arrive at
\begin{equation}
    \mathcal{F} \equiv f_\alpha  \frac{S}{I} = 1+ \frac{1}{G_1} \underbrace{\left[\alpha \frac{\beta^2G_1}{24} + \frac{S_2\beta}{4} - \frac{G_3}{6}\right]}_{= C_{\rm neq}} V^2,
\end{equation}
where we define the TUR-breaking factor $C_{\rm neq}$. Namely, if $C_{\rm neq}<0$ the TUR is violated for small voltages whereas it is fulfilled for $C_{\rm neq}\geq 0$. Notice that the quantum TUR introduces a new term into the expansion which is proportional to the first moment of the current. 

In order to demonstrate the role of the charge in a transport process, we make use of full counting statistics (see Ref.~\cite{Levitov1993,Nazarov1999,Cuevas2003}). We assume non-interacting transport which can be described by a scattering matrix. In particular, the corresponding cumulant generating functional then corresponds to a multinomial charge distribution and the general current and zero-frequency noise can be expressed by \cite{Cuevas2003,Cuevas2003a,Ohnmacht2023}
\begin{align}
    I &= \frac{1}{h}\int dE \sum_n n p_n, \\
    S &= \frac{1}{h}\int dE \left[\sum_n n^2 p_n - \left(\sum_n n p_n\right)^2\right],
\end{align}
with the charge resolved probabilities $p_n = p_n(E,V,\beta)$ which depend on energy, voltage, inverse temperature and on the \textit{charge value} $n$. Namely, $p_{\pm 1}$ corresponds to the probability of quasiparticle tunneling transferring one charge, whereas $p_{\pm 2}$ corresponds to transport mechanisms transferring two charges, like Andreev reflection. As in Ref.~\cite{Ohnmacht}, we can expand the probabilities in voltage $p_n = p_n^{(0)}+p_n^{(1)}V + p_n^{(2)}V^2/2 +...$ and use the detailed balance conditions $p_{-n}= p_n \exp(-n \beta V)$ to express the moments of current and noise through the moments of the probabilities $p_n^{(0)}$. We obtain a similar result as in Ref.~\cite{Ohnmacht}
\begin{equation}
    \frac{24C_{\rm neq}}{\beta^3} = \int \frac{dE}{h} \left[\sum_{n>0}(\alpha+2n^2)n^2 p_n^{(0)}-12\left(\sum_{n>0}n^2 p_n^{(0)}\right)^2\right].
\end{equation}
In the following, we want to investigate how to minimize this coefficient, in order to establish a corresponding TUR. We note that this function is concave in the space of probabilities $\{p_n^{(0)}\}_n$, meaning that its minimum value is obtained at the edges of the parameter space. As it holds that the sum of probabilities is upper bounded $\sum_n p_n^{(0)} \leq 1$, the function is minimized if the highest \textit{charge} value probability is maximal $p_{N}^{(0)} \neq 0$ and others are zero. Therefore, in order to address the absolute limits of the quantum TUR in systems with transport mechanisms of higher charge, we can assume that each transport channel except the highest charge value one are zero, meaning that $p_N^{(0)}\neq 0$ and $p_{n<N}^{(0)} = 0$. We therefore arrive at the following formula for the TUR-breaking factor
\begin{align}\label{Eq2}
   C_{\rm neq}  = \frac{\beta^3}{24}N^2(\alpha+2N^2) \int dE~p_N^{(0)} \left(1- \frac{12N^2}{\alpha +2N^2}p_N^{(0)}\right),
\end{align}
and it is clear that the quantum TUR ($\alpha = 1$) is more difficult to break than the classical TUR ($\alpha = 0$). 

Let us start the analysis of Eq.~\ref{Eq2} by recapping the classical TUR ($\alpha = 0$). For conventional quasiparticle transport ($N = 1$) according to Landauer, we have $p_1^{(0)} = \mathcal{T} f(1-f)$ with the transmission function $\mathcal{T} = \mathcal{T}(E)$ and the Fermi function $f = f(E)$ while $p_{n>1}^{(0)} = 0$. Thus, we retrieve the results established in Ref.~\cite{PhysRevB.98.155438}, namely in the case of a uniform transmission, the classical TUR cannot be broken. In the case when the transmission is not uniform, but rather peaked at a certain resonance value $E_0$, and temperatures are large compared to the respective broadening scale of the transmission ($\Gamma \beta \ll 1$), the contribution to the integral in Eq.~\ref{Eq2} comes from energies close to $E_0$ and is proportional to \cite{PhysRevB.98.155438}
\begin{equation}\label{pppd}
    C_{\rm neq} \propto \left[{\tau}_1 -6f(E_0)(1-f(E_0)){\tau}_2 \right],
\end{equation}
with ${\tau}_j \equiv \int dE\ \mathcal{T}^j(E)$. Requiring $C_{\rm neq} <0 $ and using that $f(1-f)\leq 1/4$ results in the condition
\begin{equation}\label{Eq3}
    \frac{\tau_2}{\tau_1}> \frac{2}{3},
\end{equation}
meaning that the classical TUR can only be violated by a non-uniform transmission function. 

Now, let us assume a Landauer type probability for a higher charge value transport mechanism, namely $p_N^{(0)} = \mathcal{T} f(1-f)$, where we assumed that the zero-voltage probability is thermally activated and thus proportional to $f(1-f)$, and $\mathcal{T} = \mathcal{T}(E)$ is an energy dependent \textit{effective} transmission function with $0 \leq \mathcal{T} \leq 1$, while $p_{n<N}^{(0)} = 0$. Noticing that for $\alpha = 0$, the prefactor reads $12N^2/(\alpha+2N^2)= 6$, we realize that the TUR-breaking condition for the classical TUR in Eq.~\ref{Eq3} is \textit{independent} on the charge value $N$. Thus, a large violation of the classical TUR is solely attributed to the energy dependence of the effective transmission $\mathcal{T}(E)$. In the case of a normal metal-superconductor junction, the effective transmission of the Andreev reflection ($p_2^{(0)}$) is effectively boxed-shaped which results in large violations of the classical TUR \cite{PhysRevB.104.045424,Ohnmacht}. 

Turning to the quantum TUR, we again assume an isolated charge transport mechanism of charge value $N$ with $p_N^{(0)} = \mathcal{T} f(1-f)$ with an effective transmission function peaked at a certain energy while $p_{n < N}^{(0)} = 0$. One can show, following the steps leading to Eq.~\ref{pppd}, that 
\begin{equation}\label{abc}
    C_{\rm neq} \propto \left[\tau_1 - 12N^2/(\alpha+2N^2)f(E_0)(1-f(E_0))\tau_2\right]
\end{equation}
and the TUR-breaking condition now reads
\begin{equation}
    \frac{\tau_2}{\tau_1} > \frac{2}{3}\left(1+\frac{\alpha}{2N^2}\right),
\end{equation}
where one easily observes that the quantum TUR ($\alpha = 1$) provides a tighter bound than the classical TUR ($\alpha = 0$). More importantly, the condition now depends on the charge value $N$. In particular, for quasiparticle tunneling in normal metal junctions where $N = 1$, we obtain $\tau_2>\tau_1 $ for the quantum TUR ($\alpha = 1$), which is never fulfilled, meaning that a one-particle transport process cannot break the quantum TUR, in accordance with Ref.~\cite{quantumTUR}. 

However, for a charge transport mechanism of higher charge value like Andreev reflection, it holds that $N = 2$, so we obtain the condition $\tau_2/\tau_1 > 3/4$, which can be achieved. It is seen that the charge value heavily impacts the validity of the quantum TUR. Note that we have not made any assumptions on the origin of these probabilities so far (we have not invoked the existence of superconductivity). In fact, despite the probability of the charge-2 transport process following an effective Landauer type behavior [$p_2^{(0)} = \mathcal{T}f(1-f)$], the quantum TUR can be broken solely because of the higher charge value. For large charge values $N \gg 1$, which would correspond to very high orders of multiple Andreev reflections, the quantum TUR converges to the classical TUR, meaning that the quantum TUR is easily broken for these systems. 
\begin{figure*}
    \centering
    \includegraphics[width=1\linewidth]{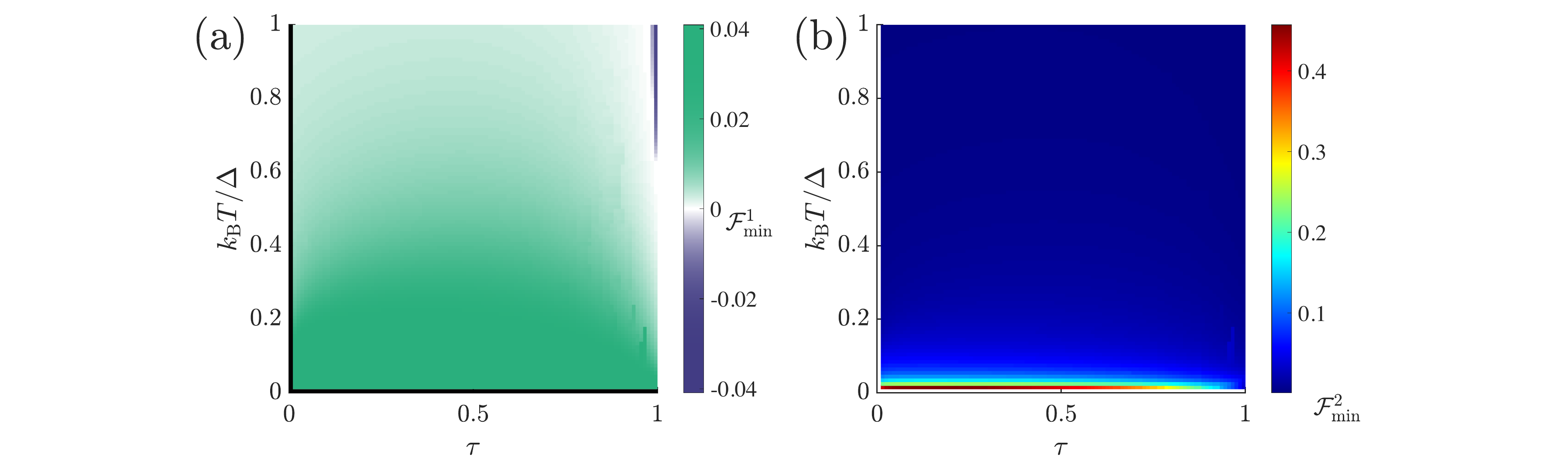}
    \caption{Investigation of the charge dependent quantum TUR for the voltage-dependent current and noise in a normal metal-superconductor point contact with one channel of energy-independent transmission $\tau$ with temperature $k_{\rm B}T/\Delta(T)$ ($\Delta \equiv \Delta(T)$ is temperature dependent superconducting gap). (a) The minimal TUR-breaking coefficient for the quantum TUR [$\mathcal{N}=1$ in Eq.~\ref{nqTUR}] for different temperatures $k_{\rm B}T$ and bare transmissions $\tau$. It is seen that the (charge-1) quantum TUR is broken for large temperatures and transmissions. (b) The minimal TUR-breaking coefficient for the charge-2 quantum TUR [$\mathcal{N}=2$ in Eq.~\ref{nqTUR}] as in (a). It is seen that the coefficient is strictly positive.}
    \label{Fig1}
\end{figure*}

\textit{Charge quantum TUR} --- In the following, we establish a new type of quantum TUR which depends on the charge value $\mathcal{N}$, namely we define the charge-$\mathcal{N}$ quantum TUR by
\begin{equation}\label{nqTUR}
    \sinh(\beta \mathcal{N} V/2) \frac{S}{\mathcal{N}I} \geq 1,
\end{equation}
which for $\mathcal{N} = 2$ is the result in Ref.~\cite{natcomm} which is derived for a superconducting hybrid system when only Andreev reflection is present. First, we establish that transport of charge values higher than one (like Andreev reflection in normal metal-superconducting junctions) obeys this TUR for $\mathcal{N} =2$ and breaks it for $\mathcal{N}=1$. Namely, upon introducing the charge value $\mathcal{N}$, the coefficient $C_{\rm neq}$ only acquires an additional factor
\begin{align}\label{conv}
    \frac{24C_{\rm neq}}{\beta^3} = \int \frac{dE}{h} \Bigg [&\sum_{n>0}(\alpha \mathcal{N}^2+2n^2)n^2 p_n^{(0)} \nonumber \\
    &-12\left(\sum_{n>0}n^2 p_n^{(0)}\right)^2 \Bigg ],
\end{align}
where we see that by considering the classical TUR ($\alpha = 0$), the charge value $\mathcal{N}$ drops out. In particular, this modification still retains $C_{\rm neq}$ to be concave and it is still minimized on the edges of the parameter space, namely if we choose the charge value $\mathcal{N}$ in Eq.~\ref{nqTUR} to be equal (or larger) to the highest charge value of transport processes $ N \leq \mathcal{N}$, the coefficient is minimized for $p_N^{(0)} \neq 0$ and all other probabilities set to zero. Hence, by assuming $p_N^{(0)} = \mathcal{T} f(1-f)$ with a peaked transmission function we obtain a modified version of Eq.~\ref{abc} with $\alpha \to \alpha \mathcal{N}^2$ and consequently arrive at
\begin{equation}\label{nqTUR2}
    \frac{\tau_2}{\tau_1} >\frac{2}{3}\Big( 1+\underbrace{\frac{\alpha \mathcal{N}^2}{2N^2}}_{\geq 1/2}\Big) \geq 1,
\end{equation}
meaning that the charge-$\mathcal{N}$ quantum TUR is always fulfilled if the maximal charge value of transport is smaller or equal to it, $N \leq \mathcal{N}$.

Whereas the above considerations have been done in the limit of small voltages and high temperatures ($\Gamma \beta \ll 1$), we can evaluate the charge quantum TUR from the full current and noise numerically for a normal metal-superconductor point contact. We follow Ref.~\cite{Ohnmacht} and define the charge dependent TUR-breaking coefficient
\begin{equation}\label{fn}
    \mathcal{F}^{\mathcal{N}} = \frac{S}{\mathcal{N}I}-\sinh(\beta \mathcal{N} V/2)^{-1}, 
\end{equation}
which is positive (or zero) if the charge-$\mathcal{N}$ quantum TUR in Eq.~\ref{nqTUR} is fulfilled and negative if the inequality is violated. In the following, we will compute this TUR-breaking coefficient for the case of $\mathcal{N} = 1$ (quantum TUR) and $\mathcal{N} = 2$ (charge-2 quantum TUR). In particular, we plot the minimal value of $\mathcal{F}^\mathcal{N}$ in Eq.~\ref{fn} for the voltage for each pair of temperature and transmission 
\begin{equation}\label{min}
    F_{\rm min}^{\mathcal{N}}(\beta,\tau) = {\rm min}_{V} F^{\mathcal{N}}(\beta,\tau,V). 
\end{equation}
In the following, we analyze the voltage-dependent current and noise in a normal metal-superconductor point contact with one channel of constant transmission $\tau$ at temperatures $k_{\rm B}T/\Delta$, where $\Delta$ is the superconducting gap, utilizing the methods in Ref.~\cite{Ohnmacht}. Notice that this transmission is not the same as the effective transmission used for the probabilities before. This \textit{bare} transmission describes the transmission probability of a particle traversing the scattering region between the normal metal and superconductor. In Fig.~\ref{Fig1}(a,b), we show the minimal value of the TUR-breaking coefficient $\mathcal{F}^1_{\rm min}$ and $\mathcal{F}^2_{\rm min}$ (see Eq.~\ref{fn}). It is seen that the (charge-1) quantum TUR is easily broken for sufficiently high temperatures and transmissions [see panel (a)] whereas the charge-2 quantum TUR is always fulfilled [see panel (b)]. This shows that the relation proposed in Ref.~\cite{natcomm} also holds in the presence of Andreev reflection \textit{and} quasiparticle tunneling. We attribute this fact to the concave nature of Eq.~\ref{conv}, which results in the function to be minimized at edges of the available parameter space.

\textit{Higher order charge mechanisms} --- In the following, we want to study higher order charge transport mechanisms like multiple Andreev reflections in superconductor-superconductor junctions. In particular, the charge numbers can heavily exceed the previously studied values of ${N} = 1,2$. We recall, that in order to minimize the charge-$\mathcal{N}$ quantum TUR, the highest charge value transport probability is non-zero, whereas all others are zero. Then, the condition for breaking charge-$\mathcal{N}$ quantum TUR is given by Eq.~\ref{nqTUR2} for arbitrary $\mathcal{N}$ and $N$. Note that if there exists a highest charge value transport mechanism of charge $N$, the charge-$\mathcal{N}$ quantum TUR cannot be broken if $N \leq \mathcal{N}$. In the case of $N = \mathcal{N}$, the TUR results in $\tau_2>\tau_1$ which cannot be fulfilled. However, in the case of unity transmission, the maximal order of multiple Andreev reflection is, in principle, unbounded. Namely, the $N$th order multiple Andreev reflection has an onset voltage of $V = 2\Delta/N$ \cite{Cuevas2003,Cuevas2003a}. Thus, for each value of $\mathcal{N}$, we will always find a process of charge value $N>\mathcal{N}$ by lowering the voltage ($N \approx 2\Delta/V$) resulting in the possibility of breaking the TUR. We infer that in the case of countably infinite charge values being present in transport, we cannot establish a corresponding charge quantum TUR. We conclude that such scenarios correspond to the ultimate limit of thermodynamic tradeoff between current-noise ratio and entropy production. Namely, the bound on the right hand side of Eq.~\ref{nqTUR2} goes to infinity for $\mathcal{N} \to \infty$.
\begin{figure*}
    \centering
    \includegraphics[width=1\linewidth]{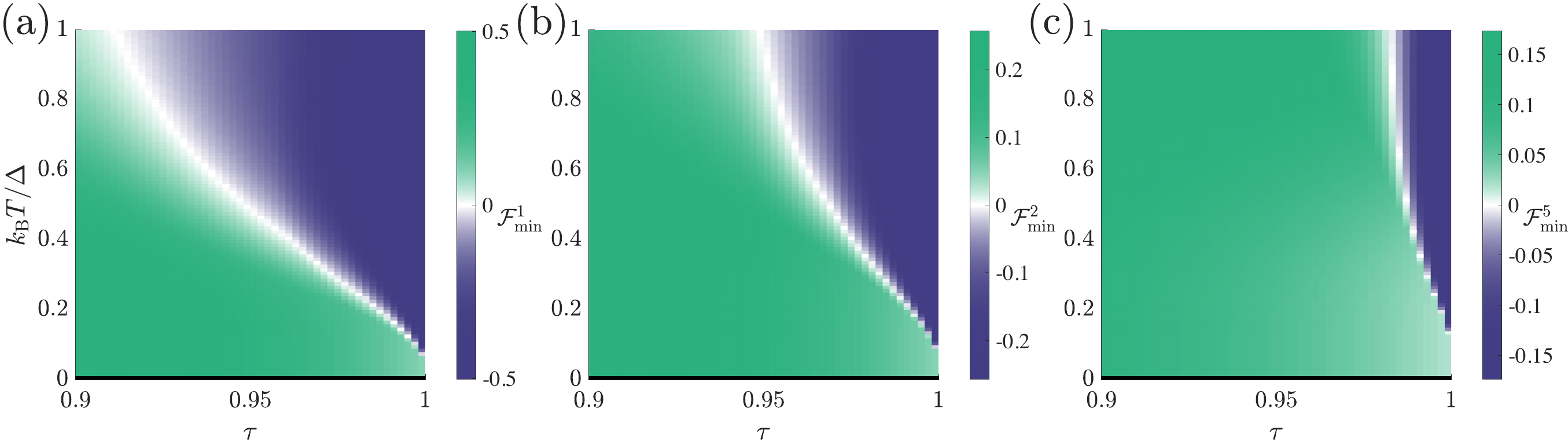}
    \caption{Investigation of the charge dependent quantum TUR for the voltage-dependent current and noise in a superconductor-superconductor point contact with one channel of energy-independent transmission $\tau$ with temperature $k_{\rm B}T/\Delta$ ($\Delta$ is superconducting gap). The minimal TUR-breaking coefficient $F_{\rm min}^{\mathcal{N}}$ for the charge-$\mathcal{N}$ quantum TUR [Eq.~\ref{nqTUR}] for $\mathcal{N} = 1$ (quantum TUR) in panel (a), $\mathcal{N} = 2$ (charge-2 quantum TUR) in panel (b) and $\mathcal{N} = 5$ (charge-5 quantum TUR) in panel (c). It is seen that the larger the charge value $\mathcal{N}$, the smaller the region where the respective TUR is broken. }
    \label{Fig2}
\end{figure*}

Nonetheless, the previous considerations enable us to quantify the degree of TUR violations in such scenarios in a systematic way. Namely, by increasing the charge value $\mathcal{N}$ of the charge quantum TUR, the bound gets increasingly more tight. In the following we consider the validity of different charge quantum TURs for transport in a superconductor-superconductor point contact with one channel of constant transmission $\tau$ by calculating the minimal TUR-breaking coefficient exactly as for Fig.~\ref{Fig1}. Namely, the full current and noise follow from the same calculations as in Ref.~\cite{Ohnmacht}. In Fig.~\ref{Fig2}(a,b,c), we illustrate the minimal TUR-breaking coefficient $F_{\rm min}^{\mathcal{N}}$ (see Eq.~\ref{fn} and Eq.~\ref{min}) for the charge-$\mathcal{N}$ quantum TUR for $\mathcal{N} = 1,2,5$ (see Eq.~\ref{nqTUR}), which corresponds to the quantum TUR, the charge-2 quantum TUR and a high charge value quantum TUR. We find that for each of these three cases, there is a combination of transmission and temperature which breaks the TUR. In particular, the breaking seems to always be present for the case of unity transmission and high enough temperatures, which relates to the current never entering the linear regime in these cases \cite{Cuevas2003,Cuevas2003a}.

\textit{Conclusion} --- We have demonstrated analytically that the highest charge value in transport dictates the corresponding charge quantum TUR. Furthermore, a charge value which is larger than one can generally lead to violations of the quantum TUR if the process is resonant. By utilizing the framework of full counting statistics, we show that the breaking is rooted solely in the charge of the transport process as the framework of full counting statistics does not invoke the existence of superconductivity. We have also shown the role of charge in the case of transport in normal metal-superconducting junctions by checking the validity of different charge quantum TURs. We find that transport in normal metal-superconducting junctions easily breaks the (charge-1) quantum TUR whereas the charge-2 quantum TUR is never broken. 

Finally, we have shown that higher order transport processes, in particular multiple Andreev reflections cannot be fundamentally bound by these charge quantum TURs, as the charge number of these processes is unbounded in ballistic scenarios ($\tau = 1$). Nonetheless, these charge-$\mathcal{N}$ quantum TURs establish a way on measuring the degree to which TUR is broken, as increasing the charge number $\mathcal{N}$ makes the bound progressively more tight.

We want to stress that we have assumed a cumulant generating function corresponding to a multinomial charge distribution. There are numerous other types of charge distributions, which would result in different versions of corresponding TURs, for example for fractional charges or super-Poissonian noise in the presence of interaction \cite{Levitov2004,Belzig2005}. Furthermore, higher orders of charge transport might eventually arise from sources other than the superconductivity, such as for microwaves resulting in photon-assisted tunneling \cite{Vanevic2007}. However, as the thermodynamic affinity does not correspond to the voltage, a dedicated analysis is required to properly quantify TURs in these cases.

D.C.O. and W.B. acknowledge support by the Deutsche Forschungsgemeinschaft (DFG; German Research Foundation) via SFB 1432 (Project No. 425217212). J.C.C.\ thanks the Spanish Ministry of Science and Innovation (contract no.\ PID2024-157536NB-C22) and the ``Maria de Maeztu" Programme for Units of Excellence in R\&D (CEX2023-001316-M). 

\bibliographystyle{apsrev4-2}
\bibliography{diebib}

\end{document}